\documentclass{article}

\usepackage{PRIMEarxiv}

\usepackage[utf8]{inputenc} 
\usepackage[T1]{fontenc}    
\usepackage{hyperref}       
\usepackage{url}            
\usepackage{booktabs}       
\usepackage{amsfonts}       
\usepackage{nicefrac}       
\usepackage{microtype}      
\usepackage{lipsum}
\usepackage{fancyhdr}       
\usepackage{graphicx}       
\graphicspath{{media/}}     
\usepackage{amsmath,amssymb,amsfonts}
\title{Entanglement Distribution and Quantum Teleportation in Higher Dimension over the Superposition of Causal Orders of Quantum Channels
\thanks{\textit{This material is based upon work supported by Science Foundation Ireland (SFI) and is co-funded under the European Regional Development Fund under Grant Numbers 13/RC/2077 and 13/RC/2077-P2. \\
\copyright 2023 IEEE. Personal use of this material is permitted. Permission from IEEE must be obtained for all other uses, in any current or future media, including reprinting/republishing this material for advertising or promotional purposes, creating new collective works, for resale or redistribution to servers or lists, or reuse of any copyrighted component of this work in other works.}} 
}

\author{\uppercase{I. Dey} and \uppercase{N. Marchetti}}

\begin{document}
\maketitle

\begin{abstract}
Multiple photonic degrees of freedom can be explored to generate high-dimensional quantum states; commonly referred to as `qudits'. Qudits offer several advantages for quantum communications, including higher information capacity, noise resilience and data throughput, and lower information loss over different propagation mediums (free space, optical fibre, underwater) as compared to conventional qubits based communication system. However, qudits have been little exploited in literature, owing to their difficulty in transmission and detection. In this paper, for the first time, we develop and formulate the theoretical framework for transmission of classical information through entanglement distribution of qudits over two quantum channels in superposition of alternative causal order. For the first time we i) engineer quantum switch operation for 2-qudit systems and ii) formulate theoretical system model for entanglement distribution of qudits via quantum switch. Results show that entanglement distribution of a qudit provides a considerable gain in fidelity even with increase in noise.
\end{abstract}


\section{Introduction}\label{S1}
{C}{ommunication} networks are gradually emerging as critical infrastructure for all walks of life. Presently, the choicest mode of communication for any network is combination of radio frequency (RF) wireless links for fronthaul and optical fibre links for backhaul connection. However, accommodating increasing number of users and applications in the RF band will be impossible in terms of carrier frequency and bandwidth. On the other hand, optical fibre links impose high latency over long-distance connections. Quantum communication networks, in future, will be the most promising alternative that can solve problems with wireless and fibre links \cite{a1}.

Quantum communication between any two points is viable through a property called `shared entanglement'. Shared entanglement can be accomplished by distributing an Einstein-Podolsky-Rosen (EPR) pair between two classically or quantumly operated nodes communicating through a quantum channel \cite{a2}. Entangled qubits are distributed between two nodes with high fidelity. Since quantum states cannot be copied or amplified, success of quantum communication over long distances strongly depends on the use of one or more quantum repeaters that forward information by entanglement distribution \cite{a3}.

Quantum particles experiencing two quantum processes in superposition of alternative orders (i.e. which channel is traversed before the other is not imperative) are capable of conveying more information than when travelling through each process of arbitrary order separately and consecutively \cite{x1}. For example, in normal condition, no quantum information can be transmitted over either of the pair of dephasing channels \cite{x2}. However, a qubit travelling through the superposition of two dephasing channels of alternative orders can reach the receiver error-free with 25\% probability. The success of an error-free transmission can only be validated by the outcome of a measurement on the order of the qubit \cite{x3}, where the order depends on the state of the quantum degree of freedom. Similarly, for partially dephasing channels, quantum capacity of two superposed channels is larger than that achievable over individual channels.

The travelling quantum particles can arrive intact at the receiver with a finite probability of success. This advantage can be capitalized on through the implementation of a quantum `switch' in a quantum teleportation chain for entanglement distribution, a phenomenon by which a pair of maximally entangled qubits must be generated and shared between the transmitter and the receiver \cite{x4}. In a quantum switch, the transmitted quantum particle experiences two quantum channels $\mathcal{A}$ and $\mathcal{B}$ in a superposition of different orders, {}{$\mathcal{A} \to \mathcal{B}$ and $\mathcal{B} \to \mathcal{A}$}. The order depends on the state of a quantum degree of freedom, which can be two-dimensional for qubits.

This phenomenon can be leveraged to create a perfect noiseless communication channel through quantum superposition of two noisy channels of alternative orders. As a result, fidelity of a teleported qubit via an ordinary channel can be enhanced by using a maximally entangled state as a quantum resource in a superposition of alternative orders \cite{b1}. Moreover, quantum entanglement is easily affected by noise resulting in degeneration  of the transmitted information, an effect that can be significantly ameliorated by the application of quantum switch enabling superposition of noisy quantum channels of different causal orders \cite{b2}.

Qubits can only exploit two photonic degrees of freedom and are therefore limited in channel capacity, data throughput and noise resilience, thereby lowering the possibility of breaking the Shannon's limit on achievable capacity. This has spurred a recent interest in teleporting qudits as units for information exchange with arbitrary dimensions exploiting $d$-degrees of freedom \cite{b3}. Teleportation of photonic qudits increases the quantum information sent per carrier photon. Teleportation of qudits can be achieved either by preparing $d$ additional photons in a highly entangled state \cite{b4} or by transcribing qudit encoded on a single photon to $d$ qubits carried by light modes which propagate along different optical paths \cite{b5}. Moreover, qudits can carry information over a plethora of media ranging from free space optical links to multi-core and multi-mode optical fibres, and peer-to-peer underwater acoustic communications \cite{x5}. 

Complementary to the preliminary efforts of the quantum physics \cite{b6} and quantum communication engineering community \cite{b7} in exploiting quantum switch for entanglement distribution of 2-qubits system, we investigate the possibility of employing quantum switch for entanglement distribution of qudits teleportation system. To achieve this, we formulate the theoretical system model for quantum switch for entanglement distribution of 2-qudit systems. We also derive closed-form expressions for connecting the teleported qudit received by Bob to the degradation suffered by the entangled pair during the distribution process.

The rest of the paper is organized as below. Section~\ref{S2} introduces the concept and mathematical expressions for the functionality of a quantum switch for qudits. Section~\ref{S3} develops the basic framework for quantum teleportation of qudits. Section~\ref{S4} follows up with the introduction to different circuit configurations possible for realizing a qudit switch and the theoretical framework for entanglement distribution of qudits. Finally, we develop the mathematical formulation for teleportation of qudits through the implementation of entanglement distribution of a 2-qudit system by means of a qudit-based quantum switch in Section~\ref{S5}. The concluding remarks are provided in Section~\ref{S6}.

\section{Quantum Switch - Concept for Qudits}\label{S2}

Let us consider an arbitrary qudit $|\psi\rangle = \{|k\rangle\}_{k=0}^{d-1}$ of dimension $d$ and unknown state travelling through two noisy quantum channels $\mathcal{A}$ and $\mathcal{B}$, and let us assume that $\mathcal{A}$ is the \emph{cyclic shift} channel and $\mathcal{B}$ is the \emph{cyclic clock} channel. The cyclic shift channel $\mathcal{A}$ performs a cyclic permutation of the basis states and maps the set, $\{|0\rangle, |1\rangle,\dotso, |d-1\rangle \}$ to $\{|1\rangle, |2\rangle, \dotso, |0\rangle\}$ and vice-versa with a probability $p$, leaving the qudit unaltered with probability $1 - p$, such that,
\begin{align}
\mathcal{A}(|\psi\rangle) = (1 - p)|\psi\rangle + p X_d|\psi\rangle
\label{eq1}
\end{align}
where $X_d$ denotes the generalized Pauli-X gate given in Table~\ref{tab:a}. {}{Here (\ref{eq1}) is the qudit generalization of the bit-flip channel that introduces a flip of the state of a qubit from $|0\rangle$ to $|1\rangle$ with a certain probability.}

The cyclic clock channel $\mathcal{B}$ introduces relative phase shift of $\omega = e^{2\pi i /d}$ between complex set of amplitudes with probability $q$, leaving the qudit unaltered  with probability $1 - q$, such that,
\begin{align}\mathcal{B}(|\psi\rangle) = (1 - q)|\psi\rangle + q Z_d|\psi\rangle\label{eq2}\end{align}
where $Z_d$ denotes the generalized Pauli-Z gate  given in Table~\ref{tab:a}. {}{Here, (\ref{eq2}) is the qudit generalization of the phase-flip channel that introduces a relative phase-shift between complex amplitudes.} The complex set of amplitudes are given by $\alpha_k$ such that, $|\psi\rangle = \sum_{k = 0}^{d - 1}\alpha_k |k\rangle$, where $\alpha_k \in \mathbb{C}^d, \sum_{k = 0}^{d - 1}|\alpha_k|^2 = 1$, $\mathbb{C}^d$ is the computational basis of $d$-dimensional Hilbert space, $\mathcal{H}^d$. A $n$-qudit state, in the tensor product Hilbert space is given by, $\mathcal{H}^{\otimes n} = (\mathbb{C}^d)^{\otimes n}$. 

The standard basis of $\mathcal{H}^{\otimes n}$ is the orthonormal basis given by the $d^n$ classical $n$-qudits, such that 
\[
|k_1 \dotso k_n \rangle = |k_1 \rangle \otimes \dotso |k_n \rangle
\label{eq3}
\]
where $0 \leq k_i \leq d - 1$ (for $i = 1, \dotso, n$). Since a qudit itself is an arbitrary superposition of orthonormal basis states $|k\rangle$ for $d$-dimensional quantum system, quantum switch for qudits will operate in a slightly different way with respect to quantum switch for qubits. For example, quantum circuit architectures completely described by instances of the CNOT gate cannot implement a transposition of a pair of qudits for dimension $d > 2$. Therefore, in order to design quantum switch for qudits, we need to consider the problem of generalizing the design beyond the qubit settings.

Next we introduce the concept of quantum switch for qudits where the traversing particles experience superposition of different orders, depending on the $d$ degrees of freedom. Whenever the control qudit is initialized to one of the basis states $|\xi\rangle$, such that $|\xi\rangle$ is a prime $|\xi\rangle \geq 2$, the quantum switch enables the message $m$ to experience the alternative classical trajectory $\mathcal{B} \to \mathcal{A}$ representing channel $\mathcal{B}$ being traversed before channel $\mathcal{A}$. Whenever the control qudit is initialized to a non-prime basis state, the quantum switch enables the message $m$ to experience the classical trajectory $\mathcal{A} \to \mathcal{B}$ representing channel $\mathcal{A}$ being traversed before channel $\mathcal{B}$. Whenever the control qudit is initialized to a superposition of the basis states, such as, 
\begin{align}|\varphi\rangle = \frac{1}{\sqrt{d}} \sum_{j = 0}^{d - 1}e^{2\pi i \varphi j/d}|j\rangle\label{eq4}\end{align}
where $\varphi$ is an integer from the set $\{0, \dotso, d - 1\}$ and $|\varphi\rangle$ is an eigenstate of the cyclic shift operator $X_d|\cdot\rangle$. Now, since a quantum superposition of two alternative orders of noisy channels, specifically $\mathcal{A} \to \mathcal{B}$ and $\mathcal{B} \to \mathcal{A}$, can behave as a perfect quantum communication channel \cite{x1}, we will use it to conceptualize quantum teleportation of qudits through entanglement generation and distribution via the application of quantum switch for qudits.

\begin{table*}[t]
\begin{center}
\caption{Higher dimensional Quantum Gates}\label{tab:a}
\begin{tabular}{|p{1.5cm}|p{2cm}|p{6cm}|p{6cm}|}
    \hline
\textbf{Gate} & \textbf{Symbol} & \textbf{align} & \textbf{Description} \\\hline\hline
\vspace{0.1cm} Generalized Pauli-$X$ \vspace{0.1cm} & \vspace{0.1cm} \rule{5mm}{.3pt}\framebox[1.5\width]{$X_d$}\rule{5mm}{.3pt} \vspace{0.1cm} & \vspace{0.1cm} $X_d^q|k\rangle = |k\oplus q\rangle$ \vspace{0.1cm} & \vspace{0.1cm} $q$ is the shift change in standard basis \vspace{0.1cm} \\\hline\hline
\vspace{0.1cm} Generalized Pauli-$Z$ \vspace{0.1cm} & \vspace{0.1cm} \rule{5mm}{.3pt}\framebox[1.5\width]{$Z_d$}\rule{5mm}{.3pt} \vspace{0.1cm} & \vspace{0.1cm} $Z_d^l|k\rangle = \omega^{lk}|k\rangle$ \vspace{0.1cm} & \vspace{0.1cm} $l$ is the phase change in standard basis \vspace{0.1cm} \\\hline\hline
\vspace{0.1cm} Generalized Hadamard \vspace{0.1cm} & \vspace{0.1cm} \rule{5mm}{.3pt}\framebox[1.5\width]{$H_d$}\rule{5mm}{.3pt} \vspace{0.1cm} & \vspace{0.1cm} $H_d|k\rangle = \frac{1}{\sqrt{d}} \sum_{m = 0}^{d - 1} x_m$,  $x_m = (-1)^{k \cdot m}|m\rangle$
\vspace{0.1cm} & \vspace{0.1cm} $k\cdot m = \oplus_{\rho = 0}^{n - 1}k_{\rho}m_{\rho}$ is the bitwise inner product  \vspace{0.1cm} \\\hline\hline
\vspace{0.1cm} CNOT Right-Shift \vspace{0.1cm} & \vspace{0.1cm} \rule{3mm}{.3pt}\framebox[1.5\width]{$R_{c_d}$}\rule{3mm}{.3pt} \vspace{0.1cm} & \vspace{0.1cm} ${R_c}_d(|k\rangle \otimes |x\rangle) = |k\rangle \otimes |x \oplus k\rangle$ \vspace{0.1cm} & \vspace{0.1cm} $\oplus$ denotes modulo-$d$ addition and $|k\rangle \otimes |x\rangle$ is the standard basis state of $\mathcal{H}_{\mathcal{A}} \otimes \mathcal{H}_{\mathcal{B}}$ \vspace{0.1cm} \\\hline\hline
\vspace{0.1cm} CNOT~\newline Left-Shift \vspace{0.1cm} & \vspace{0.1cm} \rule{3mm}{.3pt}\framebox[1.5\width]{$L_{c_d}$}\rule{3mm}{.3pt} \vspace{0.1cm} & \vspace{0.1cm} ${L_c}_d(|k\rangle \otimes |x\rangle) = |k\rangle \otimes |x \ominus k\rangle$ & \vspace{0.1cm} $\ominus$ denotes modulo-$d$ subtraction \vspace{0.1cm} \\\hline\hline
\vspace{0.1cm} Generalized XOR (GXOR) \vspace{0.1cm} & \vspace{0.1cm} \rule{2mm}{.3pt}\framebox[1.2\width]{$\text{GXOR}_d$}\rule{2mm}{.3pt} \vspace{0.1cm} & \vspace{0.1cm} GXOR$_d|k,l\rangle = |k, k \ominus l\rangle$ & \vspace{0.1cm} GXOR is the generalized quantum XOR-gate \vspace{0.1cm} \\\hline
\end{tabular}
\end{center}
\end{table*}

\section{Quantum Teleportation using Qudits}\label{S3}

\begin{figure*}[t]
\begin{center} 
 \includegraphics[width=0.7\linewidth]{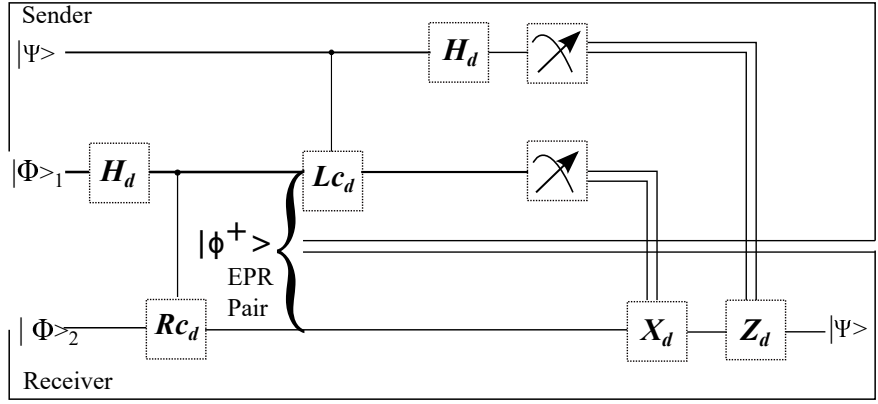}
\end{center}
\vspace*{-2mm}
\caption{Quantum Teleportation Circuit for Qudits, where $|\Psi\rangle$ is the qudit of unknown state that is to be teleported from the sender Alice to the receiver Bob, $|\Phi^{+}\rangle$ is the maximally entangled EPR pair to be generated by entangling qudits $|\Phi\rangle_1$ and $|\Phi\rangle_2$ where $|\Phi\rangle_1$ is stored at the sender and $|\Phi\rangle_2$ is stored at the receiver. The double line represents the transmission of a classical $d$-ary digit from the sender to the receiver.}
\label{figure1}
\end{figure*}

Our model of qudit teleportation protocol consists of two processes: Alice and Bob. The sender Alice possesses a qudit of unknown state 
\begin{align}|\psi\rangle_{\alpha} = \sum_{k = 0}^{d - 1}\alpha_k|k\rangle_d\label{eq5}\end{align}
that is to be teleported to Bob. This qudit is teleported using maximally entangled Einstein, Podolsky, and Rosen (EPR) pair of $|\Phi^+\rangle$ generated by entangling qudits $|\Phi\rangle_1$ and $|\Phi\rangle_2$ such that, $|\Phi\rangle_{12} = \frac{1}{\sqrt{d}} \sum_{k = 0}^{d - 1}|k\rangle_1|k\rangle_2$, through local quantum operations and classical communications {}{\cite{c5}}, as shown in Fig.~\ref{figure1}. 

Firstly, CNOT Left shift (defined in Table~\ref{tab:a}) operation is applied to qudits $|\psi\rangle_{\alpha}$ and $|\Phi^+\rangle$ followed by Hadamard operation (also defined in Table~\ref{tab:a}) applied to $|\psi\rangle_{\alpha}$. Finally, the qudits $|\psi\rangle_{\alpha}$ and $|\Phi^+\rangle$ are measured resulting in classical values ranging between 0 and $d - 1$. Using the classical values Bob can perform necessary unitary operations to recover the original state $|\psi\rangle_{\alpha}$. To begin with, Alice will make a joint von Neumann measurement of the qudit $|\psi\rangle_{\alpha}$ and the qudit $|\Phi\rangle_1$. To make this measurement, Alice will use the Bell basis of the qudits,
\begin{align} \label{eq1a}
|\Psi_{yz}\rangle = \frac{1}{\sqrt{d}} \sum_{k = 0}^{d - 1} e^{2\pi i y k/d}|k\oplus z\rangle|k\rangle
\end{align}
where $k \oplus z$ means sum of $k$ and $z$ modulo $d$, $y, z = 0, 1, \dotso, d - 1$, {}{$y$ and $z$ can take integer values between 0 and $d - 1$ and form the Bell basis for the qudits such that $|\Psi_{yz}\rangle$ form a set of basis vectors for a two-qudit systems. $|\Psi_{yz}\rangle$ can be inverted to obtain,} 
\begin{align} \label{eq1b}
|st\rangle = \frac{1}{\sqrt{d}}\sum_{y,w = 0}^{d - 1} e^{-2\pi i y k/d}\delta_{t, s\oplus w}|\Phi^{yw}\rangle.
\end{align}
The combined state of the system $|\psi\rangle_{\alpha}|\Phi\rangle_{12}$ in terms of the above Bell basis vectors for the system $|\psi\rangle_{\alpha}|\Phi\rangle_{1}$ is as follows, 
\begin{align} \label{eq1c}
|\psi\rangle_{\alpha}|\Phi\rangle_{12} = \frac{1}{d}\sum_{y,v = 0}^{d - 1}|\Psi_{yv}\rangle_{\alpha 1} U^{\dagger}_{yv}|\chi\rangle_2
\end{align}
where these unitary operators $U_{vw}$ are given by,
\begin{align} \label{eq1d}
U_{vw} = \sum_{k = 0}^{d - 1} e^{2\pi i v k/d}|k\rangle\langle k\oplus w|
\end{align}
and obey the following orthogonality conditions, $U^{\dagger}$ is the adjoint of $U$ and $U^{\dagger}$ is the conjugate transpose of $U$ in basis states, $\text{Tr}(U^{\dagger}_{vw}U_{yz}) = d~\delta_{vy}\delta_{wz}$, where $\text{Tr}$ denotes trace of a square matrix and $\delta_{vy}$ and $\delta_{wz}$ are Kronecker delta functions. {}{We introduced the indices $v$ and $w$, such that they can take any integer between 0 and $d - 1$. We introduced these indices to differentiate between the unitary operators used by Alice ($U_{yv}$) and Bob ($U_{vw}$). Alice uses the unitary operator to control the qudit state for teleportation and Bob uses a different set of operators to convert its qudit state back to the input state.}
\begin{figure*}[t]
\begin{center} 
 \includegraphics[width=0.75\linewidth]{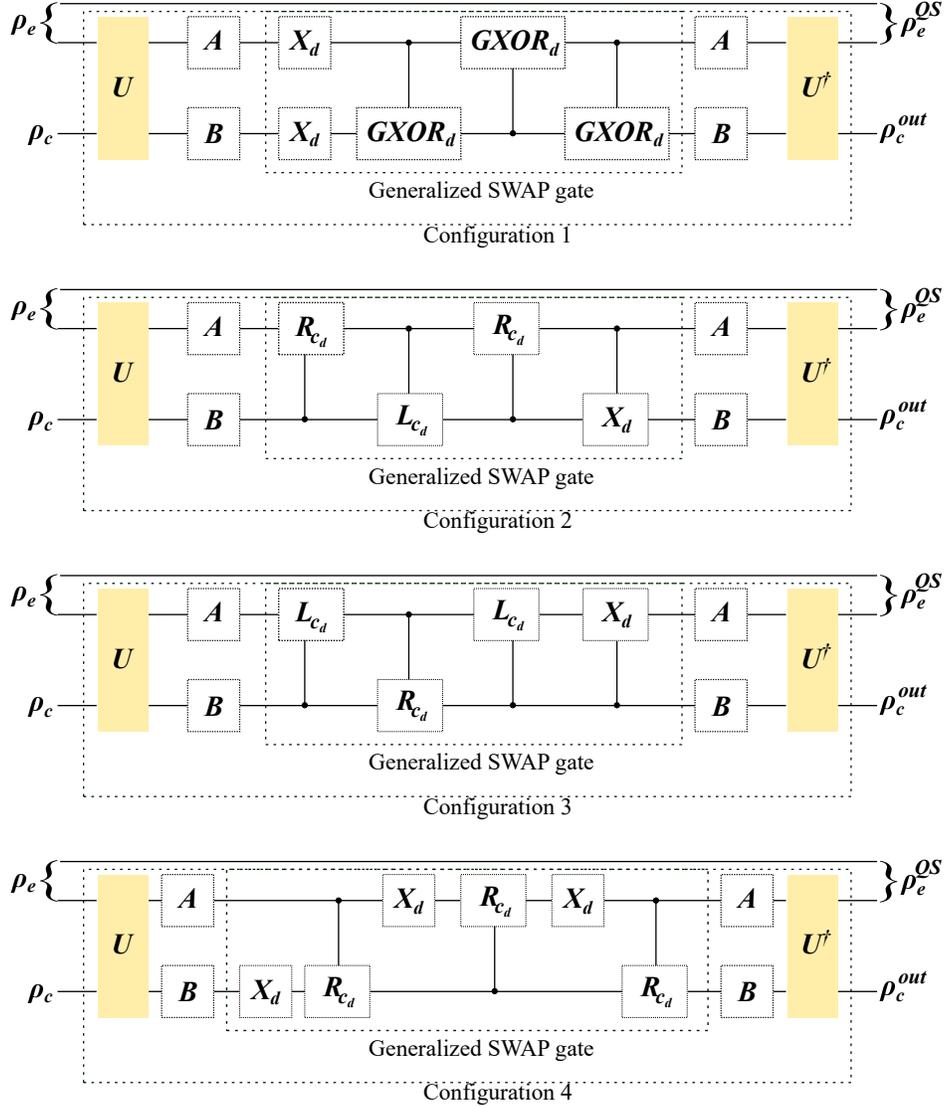}
\end{center}
\vspace*{-4mm}
\caption{Different circuit realizations for distributing and teleporting qudit entanglement where the $U$-gates control the trajectory path of the entanglement carrier from $\mathcal{A} \to \mathcal{B}$ and from $\mathcal{B} \to \mathcal{A}$ depending on the control qudit $\rho_c$. The $U^{\dagger}$-gates recombine the paths $\mathcal{A} \to \mathcal{B}$ trajectories of the entanglement carriers to generate the final density $\rho_e^{QS}$ of the EPR pair distributed between the sender and the receiver through the generalized SWAP gate for qudits.}
\label{figure2}
\end{figure*} 

In summary, Alice makes the von Neumann measurement in the Bell basis to obtain one of the possible $d^2$ results. She conveys the results of her measurement to Bob by sending $2\log_2d$ classical bits of information. After receiving this information Bob uses appropriate unitary operator $U_{vw}$ on his qudit to convert its state to that of the input state, thereby completing the standard teleportation of a qudit $|\psi\rangle_{\alpha}$ of arbitrary state over quantum channels. Though quantum teleportation with the aid of quantum entanglement looks promising, it is a very fragile resource and is easily degraded by noise \cite{b13}, resulting in loss of teleported information. However, the situation can be ameliorated by exploiting quantum superposition of different causal orders realized through a quantum switch.

\section{Entanglement Distribution of Qudits}\label{S4}

We aim to engineer entanglement distribution of qudits and thereby develop the theoretical framework for entanglement distribution of qudits via quantum switch from a communications engineer's point of view, and different circuit realizations to achieve this are given in Fig.~\ref{figure2}. 

The operator $U$ routes the entanglement carrier $|\Phi^+\rangle_2$ through either channel $\mathcal{A}$ or $\mathcal{B}$ depending on the state of $|\varphi_c\rangle$. The combination of $X_d$ and GXOR gates performs the function of routing the entanglement carrier through the other portion of the circuit through the trajectories $\mathcal{A} \to \mathcal{B}$ and $\mathcal{B} \to \mathcal{A}$ respectively (refer to Fig.~\ref{figure2}-Configuration 1). {}{The $U^{\dagger}$-gate is responsible for recombining the $\mathcal{A} \to \mathcal{B}$ trajectories and generating the final set of EPR pairs to be distributed between the transmitter and the receiver.} It is worth-mentioning here that the GXOR is the generalized quantum XOR-gate (refer to Table~\ref{tab:a}) working according to

\begin{align}\text{GXOR}_d|k,l\rangle = |k, k \ominus l\rangle.\label{eq6}\end{align}
A quantum switch for qudits can be implemented through the generalized SWAP gate operation. The generalized SWAP gate is not a single basic quantum gate for qudits; it has to be engineered using combinations of basic qudit gates. The main circuit idea for qudit SWAP, in this paper, is taken from \cite{b8} where three GXOR and two $X_d$ gates are implemented for interchanging control and target qudits between two traversing quantum channels.

Generalized SWAP gate can also be realized through several other combinations of basic higher dimensional quantum gates. A few common combinations involve CNOT Right-Shift ($R_{c_d}$), CNOT Left-Shift ($L_{c_d}$) and generalized Pauli-X ($X_d$) gates (refer to Table~\ref{tab:a}). Using these combinations, we extend our design for quantum switch in Fig.~\ref{figure2}-Configuration 1 to different implementational ways depicted in Figs.~\ref{figure2}-Configuration 2,~\ref{figure2}-Configuration 3 and ~\ref{figure2}-Configuration 4. The SWAP Gate for qudits can be implemented in many different ways depending on the requirement. The design can involve from at least 4 gates to 6 gates or more. The qudit GXOR gate can be combined with qudit Pauli $X$-gates in a cyclic fashion to generate the SWAP operation \cite{b9, b10, b11}. On the other hand, the qudit CNOT gates can also be combined with Pauli-$X$ gates in a cyclic fashion without the involvement of any GXOR gates. In that case, they are extensions of SWAP operation for qubits to qudits. This is because, a combination of a single CNOT and a single Pauli $X$-gate is used to implement SWAP operation for qubits only. It is worth-mentioning here, that three CNOT gates can be concatenated to implement two-qubit SWAP operation, where the control and the target qubits are exchanged in the second gate \cite{b12}. However, the choice of the design of the SWAP gate will have an impact on the overall operation of the quantum switch. By comparing the quantum fidelity and channel capacity achievable through qudit teleportation over each of the proposed design for quantum SWAP gates (refer to Fig.~\ref{figure2}), we can choose the best possible combination for the design of a quantum switch. We leave this interesting design detail for our future work.

To analytically model a quantum switch, we extend the concept of switch for one-qubit system to that for a one-qudit system $\mathcal{P}(\mathcal{A}, \mathcal{B}, \rho_c)(\rho)$ as the output function of the two channels $\mathcal{A}$ and $\mathcal{B}$ along with the control state $\rho_c = |\varphi_c\rangle\langle\varphi_c|$ of the control qudit,
\begin{align} \label{eq1e}
\mathcal{P}(\mathcal{A}, \mathcal{B}, \rho_c)(\rho) = \sum_{s,t}W_{st}(\rho \otimes \rho_c)W^{\dagger}_{st}
\end{align}
where $\{W_{st}\}$ denotes the set of Kraus operators associated with the superposed channel trajectories given by,
\begin{align} \label{eq1f}
{}{W_{st}} = {}{\mathcal{A}_s\mathcal{B}_t \otimes |k\rangle\langle k|}
\end{align}
{}{with $\{\mathcal{A}_s\}$ and $\{\mathcal{B}_t\}$ denoting the Kraus operators associated with the channels $\mathcal{A}$ and $\mathcal{B}$ respectively. If the control quantum state is a qubit then,}
\begin{align} \label{eq1g}
{}{W_{st}} &= {}{\mathcal{A}_s\mathcal{B}_t \otimes |0\rangle\langle 0|_c + \mathcal{B}_t\mathcal{A}_s \otimes |1\rangle\langle 1|_c}.
\end{align}
{}{If the control quantum state is a qudit then $\rho_c = |\psi_c\rangle\langle \psi_c|$, and,}
\begin{align} 
{}{W_{st}} &= {}{\sum_{k,k' = 0}^{d - 1}\big[\mathcal{A}_s\mathcal{B}_t \otimes |k\rangle\langle k| + \mathcal{B}_t\mathcal{A}_s \otimes |k'\rangle\langle k'|\big]}.
\label{eq1h}
\end{align}
{}{The Kraus operators $W_{st}$ and $W_{st}^+$ satisfy the completeness property,}
\begin{align} \label{eq1i}
{}{\sum_{\{s,t\}|_{s,t = 0}^{d - 1}}W_{st}W_{st}^{\dagger}} &= {}{I_T \otimes I_c}.
\end{align}
{}{where $I_T$ and $I_c$ are the identity operators in the target and control systems respectively. This check of completeness suggests how the indices $s$ and $t$ allow for systematic reordering of the sums by isolating or regrouping.} It is worth-mentioning here that the entanglement-pair member $|\Phi^+\rangle_1$ is already at Alice's side and therefore does not need to go through the communication channel, while the second qudit $|\Phi^+\rangle_2$ is distributed to Bob. Therefore, the first qudit basically travels virtually through an ideal channel represented by identity matrix $I_{k \times k}$.

\begin{figure*}[t]
\begin{center} 
 \includegraphics[width=0.65\linewidth]{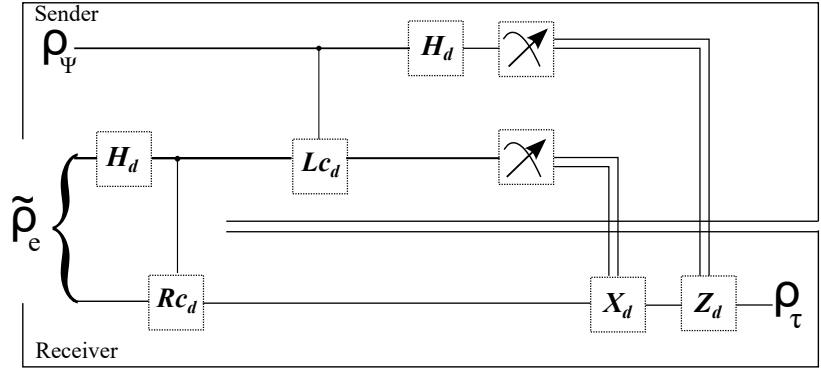}
\end{center}
\vspace*{-4mm}
\caption{Quantum Teleportation Process of qudits in terms of density matrices where $\rho_{\Psi}$, $\tilde{\rho}_e$ and $\rho_{\tau}$ denote the density matrices of the actual qudit sent by the sender, EPR pair of qudits shared between the sender and the receiver, and the actual qudit received by the receiver respectively, and the double line represents the classical trajectories.}
\label{figure3}
\end{figure*} 

Next we extend quantum switch for one qudit system to the case of a two-qudit system, represented by the density matrix $\rho_e$ that is a $d^2 \times d^2$ matrix taking into account the cyclic shift and the cyclic clock channels. In order to formulate $\rho_e$, we start by defining the Hermitian generators in the orthonormal basis $|j\rangle$ which takes on the values $j = 1, \dots, d$, given by,
\begin{align} \label{eq2c}
\Gamma_j &\equiv \frac{1}{\sqrt{j(j - 1)}}\Big(\sum_{u = 1}^{j - 1}|u\rangle\langle u| - (j - 1)|j\rangle\langle j|\Big) \nonumber\\
\Gamma_{ju}^{(+)} &\equiv \frac{1}{\sqrt{2}}\Big(|j\rangle\langle u| + |u\rangle\langle j|\Big) \nonumber\\
\Gamma_{ju}^{(-)} &\equiv \frac{-i}{\sqrt{2}}\Big(|j\rangle\langle u| - |u\rangle\langle j|\Big)
\end{align}
for $1 \leq j < u \leq d$. It is noteworthy that the generators are unitary matrices. Using orthonormality relation, we can invert the generators to give,
\begin{align} \label{eq3}
|j\rangle\langle j|&= \frac{I}{d} + \frac{1}{\sqrt{j(j - 1)}}\Big(-(j - 1)\Gamma_j + \sum_{u = j+1}^{d}\Gamma_u\Big)\nonumber\\
|j\rangle\langle u|&=  \frac{1}{\sqrt{2}}\Big(\Gamma_{ju}^{(+)} + i\Gamma_{ju}^{(-)}\Big)\nonumber\\
|u\rangle\langle j|&=  \frac{1}{\sqrt{2}}\Big(\Gamma_{ju}^{(+)} - i\Gamma_{ju}^{(-)}\Big)
\end{align}

We consider a class of two qudit states $|\Phi\rangle_1$ and $|\Phi\rangle_2$ with a maximally entangled state, which can be chosen to be $|\Phi^+\rangle = \frac{1}{\sqrt{d}}\sum_{j = 1}^d |j\rangle \otimes |j\rangle$. By assuming without any loss of generality $\rho_e$ being the $d^2 \times d^2$ density matrix associated with the EPR pair $|\Phi^+\rangle$, we can calculate it from the inverted generators to give,
\begin{align} \label{eq4}
&\rho_e = |\Phi^+\rangle\langle\Phi^+| = \frac{1}{d^2}(I \otimes I) \nonumber\\
&+ \frac{1}{d}\sum_j \Big[\Gamma_j\otimes\Gamma_j + \Gamma_{ju}^{(+)}\otimes\Gamma_{ju}^{(+)} +\Gamma_{ju}^{(-)}\otimes\Gamma_{ju}^{(-)}\Big].
\end{align}

\begin{figure*}[t]
\normalsize
\begin{align} \label{eq5}
&{}{\mathcal{P}(\mathcal{A}, \mathcal{B}, \rho_c)(\rho) = \bigg\{(1 - p)(1 - q)\rho_e + (1 - p)q \bigg[\sum_{s}\sum_{k = 0}^{d - 1}|\lambda_{sk}\rangle\langle \lambda_{sk}|\bigg]\rho_e \bigg[\sum_{s}\sum_{k = 0}^{d - 1}|\lambda_{sk}\rangle\langle \lambda_{sk}|\bigg]^{\dagger}+ p(1 - q)\bigg[\sum_{s,t}|st\rangle\langle \lambda_{st}|\bigg]}\nonumber\\
&{}{\times\rho_e\bigg[\sum_{s,t}|st\rangle\langle \lambda_{st}|\bigg]^{\dagger}\bigg\}\otimes \bigg[\frac{1}{\sqrt{d}} \sum_{j = 0}^{d - 1} \omega^{\varphi_c j}|j\rangle\langle j|\bigg] + pq\bigg[\sum_{s,t}\iota_t|st\rangle\langle \lambda_{st}|\bigg]\rho_e\bigg[\sum_{s,t}\iota_t|st\rangle\langle \lambda_{st}|\bigg]^{\dagger} \otimes \bigg[\frac{1}{\sqrt{d}} \sum_{j = 0}^{d - 1} \omega^{\varphi_c j}|j\rangle\langle j|\bigg].}
\end{align}
\hrulefill
\end{figure*}
Following the aligns for $\rho_e$, the global quantum state at the output of the switch can be calculated by exploiting the tensor product properties after some algebraic manipulation to obtain (\ref{eq5}), where the control qudit $\varphi_c$ is transformed into a mixed state of basis states, $\lambda_{sk} = s_{\oplus(d-1)}k$, $\lambda_{st} = st_{\oplus(d-1)}$, $\omega^{\varphi_c j} = e^{i 2\pi\varphi_c j/d}$ and $\iota_t = (-1)^{t \oplus(d - 1)}$, where $\oplus (d - 1)$ in subscript represents the direct sum of $(d - 1)$ copies of the fundamental variable. The global state $\mathcal{P}(\mathcal{A}, \mathcal{B}, \rho_c)(\rho_e)$ at the output of the quantum switch is a mixture of the pure states (i.e. $\langle\psi_c||\psi_c\rangle = 1$ in a complex Hilbert space) of the control qudit. Based on the Hadamard basis, if the measurement outcome is $|-\rangle$, the global state collapses with probability $pq$ to,
\begin{align} \label{eq6a}
\rho^{QS}_{e_{|-\rangle}} = \Big[\sum_{s,t}\iota_t|st\rangle\langle \lambda_{st}|\Big]\rho_e\Big[\sum_{s,t}\iota_t|st\rangle\langle \lambda_{st}|\Big]^{\dagger}.
\end{align}
The entanglement distribution in this case is a noiseless process. Bob receives the particle $|\Phi\rangle_2$ without any error if $\rho_e^{QS} = \rho_e$, thereby yielding to a noiseless entanglement distribution through the $d$-dimensional quantum switch with probability $pq$. Alternatively, the distribution can be a noisy process degraded by a noisy communication channel with a probability $(1 - pq)$. In that case, the measurement outcome is $|+\rangle$ and the global state collapses to,
\begin{align} \label{eq6b}
&\rho^{QS}_{e_{|+\rangle}} = \frac{(1-p)(1-q)\rho_e}{1-pq} + \frac{p(1-q)\Big[\sum_{s,t}|st\rangle\langle \lambda_{st}|\Big]\rho_e\Big[\sum_{s,t}|st\rangle\langle \lambda_{st}|\Big]^{\dagger}}{1-pq} \nonumber\\
&\quad + \frac{(1-p)q\Big[\sum_{s,k}|\lambda_{sk}\rangle\langle \lambda_{sk}|\Big]\rho_e\Big[\sum_{s,k}|\lambda_{sk}\rangle\langle \lambda_{sk}|\Big]^{\dagger}}{1-pq}.
\end{align}
Subsequently we can derive the expression for the density matrix $\rho_e^{QS}$ of the EPR pair distributed between Alice and Bob at the output of the quantum switch which will be equal to $\rho_e$ with probability $pq$ and otherwise as,
\begin{align} \label{eq6}
&\rho^{QS}_e = \Big((1 - p)(1 - q)\rho_e + p(1 - q)\Big[\sum_{s,t}|st\rangle\langle \lambda_{st}|\Big]\times\rho_e\Big[\sum_{s,t}|st\rangle\langle \lambda_{st}|\Big]^{\dagger} + (1 - p)q \Big[\sum_{s}\sum_{k = 0}^{d - 1}|\lambda_{sk}\rangle\langle \lambda_{sk}|\Big] \nonumber\\
&\quad\times\rho_e \Big[\sum_{s}\sum_{k = 0}^{d - 1}|\lambda_{sk}\rangle\langle \lambda_{sk}|\Big]^{\dagger}\Big)/(1 - pq).
\end{align}

With probability $pq$ for the measurement of the control qudit corresponding to the basis states, the entanglement distribution is noiseless. Bob receives the particle $|\Phi\rangle_2$ without any error if $\rho_e^{QS} = \rho_e$, while it is degraded by noise with the probability $(1 - pq)$. Since $\mathcal{A}$ and $\mathcal{B}$ denote the bit-flip and the phase-flip channels respectively, traversed in a well-defined order $\mathcal{A} \to \mathcal{B}$, the density matrix of the entangled pair at Bob's side is given by,
\begin{align} \label{eq7a}
\rho^{CT}_e &= \mathcal{B}\big[\mathcal{A}(\rho_e)\big] = \mathcal{B}\big[\sum_s \mathcal{A}_s \rho_e \mathcal{A}_s^{\dagger}\big] \nonumber\\
&= \sum_t \mathcal{B}_t \big[\sum_s \mathcal{A}_s \rho_e \mathcal{A}_s^{\dagger}\big] \mathcal{B}_t^{\dagger}.
\end{align}
\begin{figure*}[t]
\normalsize
\begin{align} \label{eq7}
\rho^{CT}_e &= (1 - p)(1 - q)\rho_e + p(1 - q)\Big[\sum_{s,t}|st\rangle\langle \lambda_{st}|\Big]\rho_e\Big[\sum_{s,t}|st\rangle\langle \lambda_{st}|\Big]^{\dagger} + (1 - p)q \Big[\sum_{s}\sum_{k = 0}^{d - 1}|\lambda_{sk}\rangle\langle \lambda_{sk}|\Big] \rho_e \nonumber\\
&\times \Big[\sum_{s}\sum_{k = 0}^{d - 1}|\lambda_{sk}\rangle\langle \lambda_{sk}|\Big]^{\dagger} + pq\Big[\sum_{s,t}\iota_t|st\rangle\langle\lambda_{st}|\Big]\rho_e\Big[\sum_{s,t}\iota_t|st\rangle\langle \lambda_{st}|\Big]^{\dagger}.
\end{align}
\hrulefill
\end{figure*}
Consequently, the expression for the density matrix $\rho_e^{CT}$ of the EPR pair distributed between Alice and Bob for classical trajectories is given by (\ref{eq7}). Using (\ref{eq5}, \ref{eq6}, \ref{eq7}), it is possible to calculate the performance gain achievable through the teleportation of qudit over that achievable through the teleportation of qubit. Also it is possible to compute the performance gain achievable through entanglement distribution via quantum switch (for qudits) over that achievable without any entanglement distribution. 

\section{Qudit Teleportation through Quantum Switch}\label{S5}

Using the expression for the density matrix of the teleported qudit at Bob's side, we can quantify the effect of noise on EPR distribution on the teleported qudit. Towards that end, let us assume $\rho_{\psi} \text{Tr}iangleq |\psi\rangle_a\langle\psi|_a$ be the $d \times d$ density matrix of the unknown pure qudit state $|\psi\rangle_a$ that Alice sends to Bob, such that,
\begin{align}
\rho_{\psi} =
\begin{bmatrix}
\rho^{11}_{\psi} & \cdots & \rho^{1d}_{\psi} \\
\vdots & \ddots & \vdots \\
\rho^{d1}_{\psi} & \cdots & \rho^{dd}_{\psi}  
\end{bmatrix}
\end{align}
Let us assume that the expression for the density matrix of the distributed EPR pair at the output of the quantum switch corrupted by noise be equal to $\tilde{\rho}_e$. If the entanglement distribution is perfect, $\tilde{\rho}_e = \rho_e$. Here $\tilde{\rho}_e$ is a $d^2 \times d^2$ matrix consisting of $d\times d$ block matrices to yield,
\begin{align}
\tilde{\rho}_e =
\begin{bmatrix}
\tilde{\rho}_{e_{11}} & \cdots & \tilde{\rho}_{e_{1d}} \\
\vdots & \ddots & \vdots \\
\tilde{\rho}_{e_{d1}} & \cdots & \tilde{\rho}_{e_{dd}}  
\end{bmatrix}
\end{align}
We also assume that the density matrix of the teleported qudit at Bob's side is equal to $\rho_{\tau}$. Using these notations, we represent the quantum teleportation process of a qudit in terms of density matrices in Fig.~\ref{figure3}.
\begin{figure}[t]
\begin{center} 
 \includegraphics[width=1\linewidth]{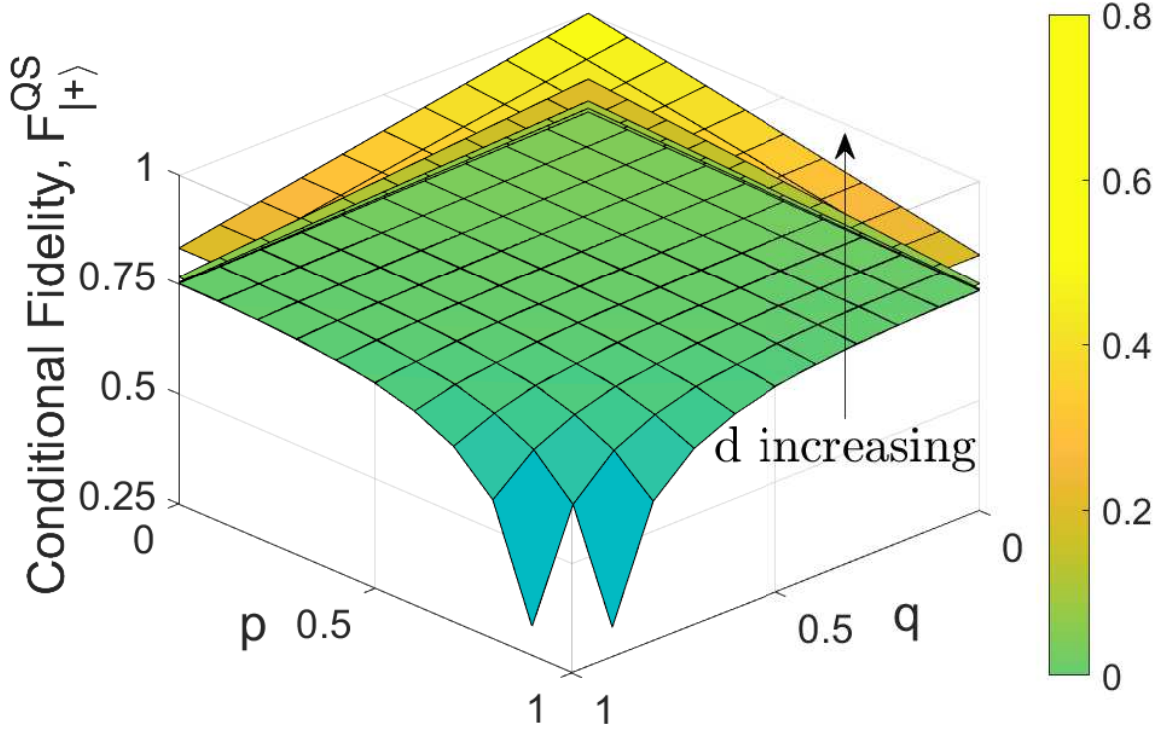}
\end{center}
\vspace*{-4mm}
\caption{Conditional fidelity of the teleported qudit as a function of the error probabilities $p$ and $q$ of the noisy bit flip channel where measurement of the control qudit $|\psi_c\rangle$ yields $|+\rangle$.}
\label{figure5}
\vspace*{-5mm}
\end{figure} 
\begin{figure}[t]
\begin{center} 
 \includegraphics[width=1\linewidth]{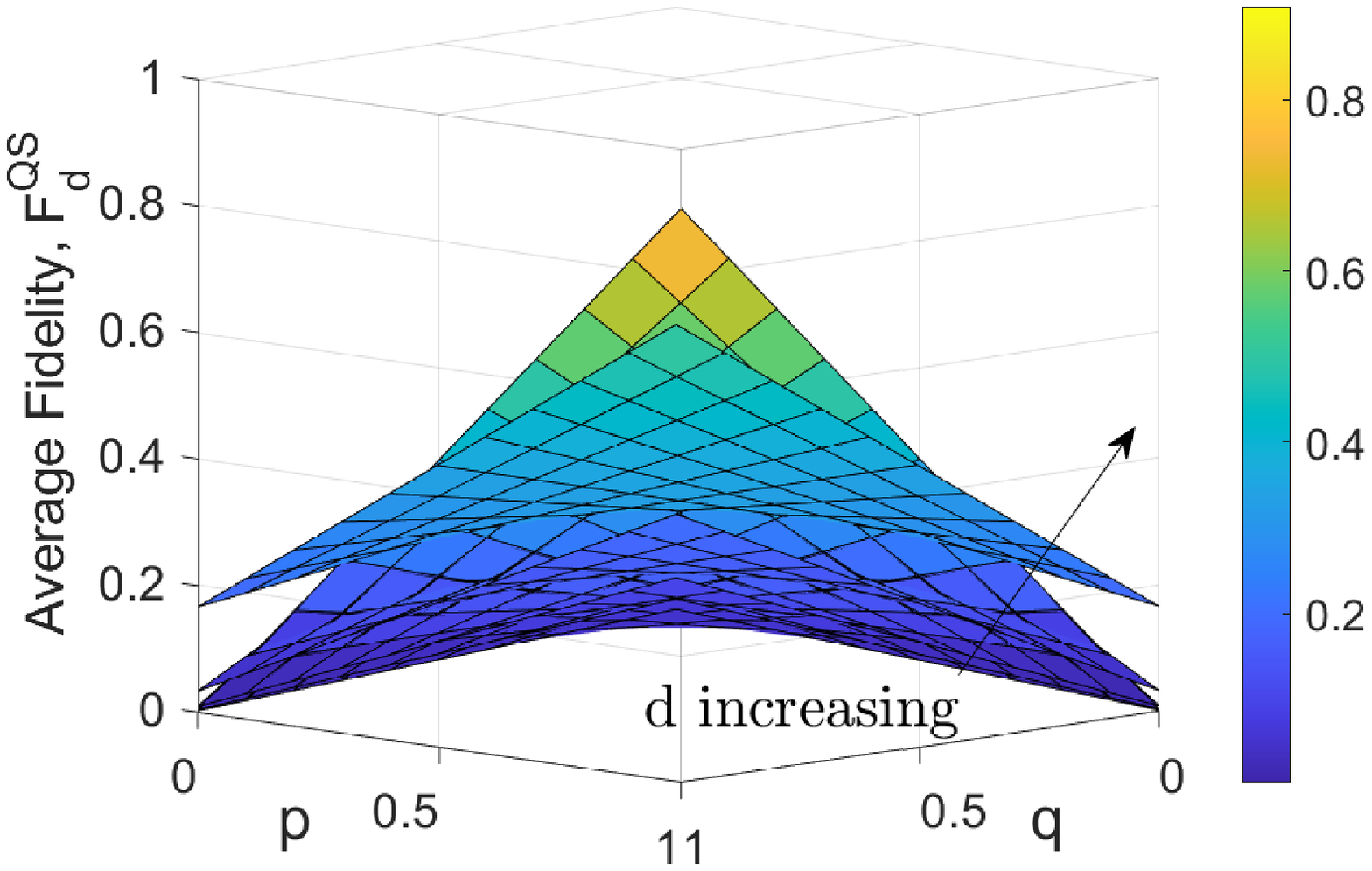}
\end{center}
\vspace*{-4mm}
\caption{Average fidelity of the teleported qudit as a function of both the error probabilities $p$ and $q$.}
\label{figure6}
\vspace*{-5mm}
\end{figure} 

Following the procedure in \cite{b7}, we first calculate $\rho_{\tau}$ if the measurement outcome is $|kk\rangle$ (for $k = 0, \dotso, d -1$) to obtain,
\begin{align}
    \rho^{|kk\rangle}_{\tau} = d \Big[\sum_{k = 1}^d \rho^{kk}_{\psi} \tilde{\rho}_{e_{kk}} + \sum_{v = 1}^j\sum_{u = j+1}^d \rho^{vu}_{\psi} \tilde{\rho}_{e_{vu}}\Big].
\end{align}
If the measurement outcome is $|vu\rangle$ (for $v = 0, \dotso, j -1$, $u = j, \dotso, d -1$) $\rho_{\tau}$ can be given by,
\begin{align}
    \rho^{|vu\rangle}_{\tau} = d \Big[\sum_{k = 1}^d \rho^{kk}_{\psi} \tilde{\rho}_{e_{kk}} - \sum_{v = 1}^j\sum_{u = j+1}^d \rho^{vu}_{\psi} \tilde{\rho}_{e_{vu}}\Big].
\end{align}
Using $\rho_{\tau}$ we can calculate the performance gain achievable through the superposition of causal orders via the quantum switch in terms of quantum fidelity $F$. The conditional fidelity $F^{QS}_{|+\rangle}$ associated with the teleported qudit at Bob's side can be calculated in terms of $\rho_{\tau}$ to obtain \cite{b7},
\begin{align}
    F^{QS}_{|+\rangle} = \text{Tr}\big[\rho_{\tau} \rho_{\psi}\big] = \frac{d - dp - dq + pq + 1}{(1 - pq)(d + 1)}
\end{align}
while the conditional fidelity $F^{QS}_{|-\rangle} = 1$ with probability $pq$, when the measurement of the control qudit corresponds to the state $|+\rangle$ and $|-\rangle$ respectively. {}{In this case, the control qudit $|\psi_c\rangle = |+\rangle\langle+|$ where $|+\rangle \equiv \frac{1}{\sqrt{d}}(|1\rangle + |2\rangle + \dotso + |d\rangle)$ or $|\psi_c\rangle = |-\rangle\langle-|$ where $|-\rangle \equiv \frac{1}{\sqrt{d}}(|1\rangle - |2\rangle - \dotso - |d\rangle)$}. The average fidelity $\bar{F}^{QS}$ of the teleported qudit on Bob's side using quantum switch can be given by,
\begin{align}
    \bar{F}^{QS} &= pq F^{QS}_{|-\rangle} + (1-pq) F^{QS}_{|+\rangle} \nonumber\\
    & = \text{Tr}\Bigg[d \Big(\sum_{k = 1}^d \rho^{kk}_{\psi} \tilde{\rho}_{e_{kk}} + \sum_{v = 1}^j\sum_{u = j+1}^d \rho^{vu}_{\psi} \tilde{\rho}_{e_{vu}}\Big) \nonumber\\
    &\times \begin{bmatrix}
\rho^{11}_{\psi} & \cdots & \rho^{1d}_{\psi} \\
\vdots & \ddots & \vdots \\
\rho^{d1}_{\psi} & \cdots & \rho^{dd}_{\psi}  
\end{bmatrix}\Bigg] 
\end{align}
which can be manipulated using mathematical operations, like matrix multiplication, integration and trace of square matrices, to obtain,
\begin{align}
 \bar{F}^{QS}= \frac{d - dp - dq + d^2pq + 1}{d + 1}
\end{align}
with $p$ and $q$ being the error probabilities of the two considered noisy channels $\mathcal{A}$ and $\mathcal{B}$. It is worth-mentioning that a fidelity of 1 corresponds to the case where the received qudit on Bob's side is identical to the original qudit and the measurement of the control qudit is equal to $|-\rangle$. The gain in performance on using quantum switch for qudit teleportation can be calculated in terms of quantum fidelity and achievable channel capacity by applying quantum switch with classical trajectories, a detailed treatise of which we leave for our future work.

In Fig.~\ref{figure5}, we plot the conditional fidelity $|F^{QS}_{|+\rangle}$ with respect to $p$ and $q$, where $|F^{QS}_{|+\rangle}$ is the fidelity corresponding to the control qudit $|+\rangle$. Fig.~\ref{figure5} demonstrates that the conditional fidelity deteriorates irreversibly with the increase in error probability $p$ of the bit flip channel and $q$ of the phase flip channel, if the control qudit $|\psi_c\rangle$ is measured to be in state $|+\rangle$ for any $p, q > 0$. This degradation in fidelity results from the degradation in entanglement and thereby in the quality of the received information. 

Fig.~\ref{figure6} portrays the average fidelity as a function of the error probabilities $p$ and $q$ of the noisy channel, as long as the control qudit $|\psi_c\rangle$ is measured to be in state $|-\rangle$. The noise in the channel causes a degradation in fidelity up to a certain threshold and then the fidelity starts increasing with increase in $p$ and $q$. This happens due to the advantage offered by the qudit switch, which makes it possible to send information even when the noise increases; a task impossible to achieve within the standard quantum Shannon theory. This improvement in average fidelity after a certain point is utilized in our design for quantum switch to yield an overall improvement in fidelity even in a very noisy quantum channel. Highest possible fidelity is achieved for the case when $p = 0$ and $q = 0$; that is the case where the channels are technically noiseless. Here the threshold value refers to the values of the error probabilities $p, q$. With further increase in $p$ and $q$ after the threshold value, average fidelity increases. This threshold is actually the limiting condition of the zero-capacity quantum channel, i.e. $p = q = 0.5$. For the values of $p$ and $q$, lower than the threshold value, fidelity decreases with increase in error probabilities. This means that the advantage in terms of fidelity obtained through employing a quantum switch is only attainable after a certain threshold for $p$ and $q$ is reached.
\begin{figure}[t]
\vspace*{-2mm}
\begin{center} 
 \includegraphics[width=1.01\linewidth]{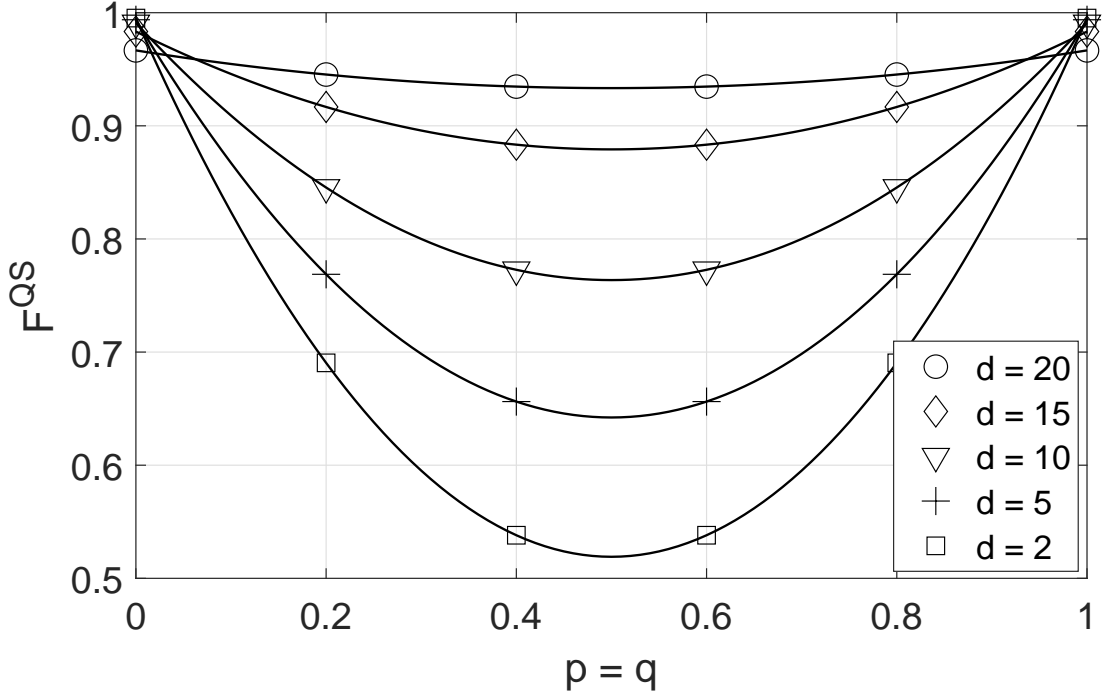}
\end{center}
\vspace*{-4mm}
\caption{Average fidelity of the teleported qudit as a function of $p$ when $q = p$ and entanglement distribution is achieved through a quantum switch.}
\label{figure4}
\vspace*{-5mm}
\end{figure} 

In Fig.~\ref{figure4}, we compare the average fidelity achievable with a quantum switch where a $d$-dimensional qudit is teleported by varying $d$, $d = 2, 5, 10, 15, 20$ as a function of the error probabilities $p$ and $q$ of the bit-flip and the phase-flip channels respectively for the condition when $q = p$. When $p = q = 0.5$ (like a 50\% erasure channel), no quantum information can be sent through any classical trajectory traversing the channels $\mathcal{A}$ and $\mathcal{B}$ \cite{b2}. Hence, achievable fidelity increases with the dimension of the quantum particle used for teleportation. Increased dimension refers to an enlarged Hilbert space and a larger Hilbert space offers the advantage of communicating with larger information capacity and higher noise resilience than traditional qubit-based quantum systems. It is evident from Fig.~\ref{figure4}, average fidelity ${\bar{F}}^{QS}$ decreases as $p$ and $q$ increase (i.e. increase in error probabilities), up to a certain threshold like $p = q = 1/2$ for the case of $d = 2$. Fidelity starts increasing after the threshold even as noise affecting the channel increases, thereby corroborating the results obtained in \cite{b7}. 

For the case when $p = q \to 0.5$ (the limiting condition of zero-capacity quantum channel), teleportation of qubit over a quantum switch offers a fidelity of approaching 0.5, i.e. $\bar{F}^{QS}_{d = 2} \to 0.5$. While as we increase $d$ to 5, fidelity approaches 0.65. If $d = 10$, $\bar{F}^{QS}_{d = 10} \to 0.78$, $d = 15$, if $\bar{F}^{QS}_{d = 15} \to 0.89$, $d = 20$, and if $\bar{F}^{QS}_{d = 20} \to 0.94$. Therefore, entanglement distribution of a qudit over a quantum switch provides a remarkable gain in fidelity of the teleported qudit at Bob's side with increase in both noise plaguing the communication channel and dimensionality of the quantum particle.

For any scenario of quantum teleportation, the EPR pair while being distributed between transmitter and receiver may lose its coherence, and become a mixed state due to interaction with the environment and noise introduced by the quantum channel. This in turn reduces the fidelity of teleportation, thereby reducing the range of states that can be accurately teleported. However, the results obtained here demonstrate that \textbf{fidelity can be improved even in presence of noise}, if a quantum switch is used. Towards this end, if entanglement is distributed between sender and receiver using a quantum switch, fidelity of teleportation can be enhanced even in presence of noise, thereby offering a wide range of states that can be accurately teleported.

\section{Conclusion}\label{S6}
In this paper, we formulated the first ever theoretical framework for implementing quantum switch for $d$-dimensional quantum teleportation system from a communication engineering point of view. With the help of generalized quantum gates, we have designed the basic scheme for teleportation of qudits and then applied it to quantum switch for entanglement distribution. In future, we will use the results obtained in this paper directly to evaluate the performance gain achievable by teleporting and distributing entanglement of qudits over that of qubits in noisy quantum channels, in terms of information carrying capacity and resilience to noise. We will also analyze the impact of designing the generalized SWAP gate (different combinations of higher dimensional quantum gates as demonstrated in Fig.~\ref{figure2}) on the overall performance of qudit teleportation in our future work.


\end{document}